\documentclass[12pt]{article}
\usepackage{iftex}
\ifPDFTeX
\pdfoutput=1
  \usepackage[utf8]{inputenc}
  \usepackage[T1]{fontenc}
  \usepackage{lmodern}
\else
  \usepackage{fontspec}
\fi
\usepackage[margin=2.5cm]{geometry}
\usepackage[english]{babel}
\usepackage{amsmath, amsfonts, amssymb}
\usepackage{graphicx}
\usepackage{dcolumn}
\usepackage{bm}
\usepackage[mathlines]{lineno}
\usepackage{upgreek}
\usepackage{soul, xcolor}
\usepackage{setspace}
\usepackage{multirow}
\usepackage{makecell}
\usepackage{rotating}
\usepackage{lipsum}
\usepackage[super,sort&compress,numbers]{natbib}
\usepackage{authblk}
\usepackage{hyperref}

\begin{document}
\onehalfspacing

\title{Ultrastrong Au-Te bonding drives disorder in monolayer ZrTe$_5$ on gold}

\author[1,2,3]{Konr\'{a}d Kandrai\textsuperscript{\textdaggerdbl}}
\author[1,2,4]{Zolt\'{a}n Tajkov\textsuperscript{\textdaggerdbl}}
\author[1]{P\'{e}ter Kun}
\author[1,3]{Levente Tapaszt\'{o}}
\author[5]{J\'{a}nos Koltai\thanks{janos.koltai@ttk.elte.hu}}
\author[1,2]{P\'{e}ter Nemes-Incze\thanks{nemes.incze.peter@ek.hun-ren.hu}}

\affil[1]{Hungarian Research Network, HUN-REN Centre for Energy Research, Institute of Technical Physics and Materials Science, 1121 Budapest, Hungary}
\affil[2]{MTA - HUN-REN EK Lendület ``Momentum'' Topology in Nanomaterials Research Group, 1121 Budapest, Hungary}
\affil[3]{Department of Physics, Institute of Physics, Budapest University of Technology and Economics, M\H{u}egyetem rkp. 3., H-1111 Budapest, Hungary}
\affil[4]{ELTE E\"{o}tv\"{o}s Lor\'{a}nd University, Department of Physics of Complex Systems, 1117 Budapest, Hungary}
\affil[5]{ELTE E\"{o}tv\"{o}s Lor\'{a}nd University, Department of Biological Physics, 1117 Budapest, Hungary}
\affil[ ]{\textsuperscript{\textdaggerdbl} These authors contributed equally.}

\date{\today}

\maketitle

\begin{abstract}
Monolayer ZrTe$_5$ is predicted to host a large-gap quantum spin Hall phase, motivating efforts to isolate single layers of the material.
Gold-assisted exfoliation produces clean monolayers of many chalcogen-terminated van der Waals crystals, but the strong Te-Au bond may compete with the bonding network of the ZrTe$_5$ layer itself.
Across tens of samples, low-temperature scanning tunneling microscopy on gold-exfoliated flakes shows a disordered monolayer surface in the overwhelming majority of cases, while thicker flakes preserve the characteristic quasi-one-dimensional chain structure.
\emph{Ab initio} calculations of the ZrTe$_5$/Au(111) interface reproduce this asymmetry and resolve its mechanism: the Te atoms facing the gold chemisorb, rupturing the weak zigzag Te-Te bonds that cross-link the ZrTe$_3$ chains, turning the discrete bond-length spectrum of the crystal into a continuous one, while the chains persist as distorted units.
Charge analysis shows the signature of covalent Te-Au bonding with a modest transfer that hole-dopes the monolayer; in the bilayer, the distortion and the doping stay confined to the layer contacting the gold, and the chemisorption and disorder each preclude the predicted quantum spin Hall phase.
The same quasi-covalent Te-Au bond may perturb the contact layer of other gold-exfoliated Te-terminated crystals, particularly those with weakly connected intralayer networks.
\end{abstract}

\section{Introduction}

ZrTe$_5$ is a layered pentatelluride predicted to host a large-gap quantum spin Hall insulator phase in the monolayer limit~\cite{weng2014transition}, with an experimental bulk gap on the order of 100~meV~\cite{wu2016evidence}.
Weak interlayer coupling along the stacking direction places the bulk close to the boundary between weak and strong three-dimensional topological phases, making the material sensitive to strain and dimensionality as knobs for topological phase control~\cite{Tajkov2022-ur,mutch2019evidence,monserrat2019unraveling,zhang2021observation}.
Isolating high-quality monolayers is a prerequisite for accessing the predicted~\cite{weng2014transition} 2D topological phase directly.
For this reason, several routes toward monolayer ZrTe$_5$ have been explored.
An Al$_2$O$_3$-assisted exfoliation yields flakes of $\sim$10~$\upmu$m lateral size~\cite{Zhuo2022-ae}, and molecular beam epitaxy on SiC/graphene produces monolayer islands with a measured band gap of $\sim$0.25~eV~\cite{Xu2024-hv}.

A complementary approach is gold-assisted exfoliation.
For chalcogen-terminated van der Waals crystals, the bottom layer adheres more strongly to a freshly prepared gold surface than to the bulk above, enabling clean separation of a single layer~\cite{Magda2015-nn,Velicky2018-wj}.
The technique delivers monolayer MoS$_2$ and related transition-metal dichalcogenides up to the centimeter scale~\cite{Velicky2018-wj,Huang2020-tw,Desai2016-kv,Li2021-kr,Yoon2026-kv}, and has been applied to tellurides including MoTe$_2$, WTe$_2$, PtTe$_2$, and PdTe$_2$~\cite{Velicky2018-wj,Huang2020-tw}.
Within this family, Te-terminated crystals bind to gold particularly strongly: experiments on self-assembled monolayers and first-principles calculations of chalcogen anchors on Au(111) find the chalcogen-gold interaction increasing from S to Se to Te~\cite{Ossowski2015-ts,Szelagowska-Kunstman2010-ss,Miranda-Rojas2016-se,Miranda-Rojas2020-ac,Garzon2026-nc}.
The affinity of gold for tellurium also shows in its mineralogy: apart from native gold, the naturally occurring compounds of gold are predominantly tellurides, calaverite (AuTe$_2$) chief among them~\cite{Okamoto1984-au,Andon1971-te}.
On surfaces, pure Te adsorption perturbs the gold itself and lifts the Au(100) surface reconstruction over large areas~\cite{Zhang2023-mf}, and ARPES on Au-Te surface alloys~\cite{Bouaziz2020-nx} and on gold-exfoliated monolayer IrTe$_2$~\cite{Asikainen2025-xf} reveals strong Te-Au hybridization with a quasi-covalent character.
Raman signatures of strain and charge doping in monolayer MoS$_2$ on gold point to a similar interaction beyond simple van der Waals adhesion~\cite{Velicky2020-lu}.
For ZrTe$_5$, where the Zr-Te bond length is close to 3~\AA~\cite{osti_1277414} and the interlayer spacing is 7.25~\AA~\cite{Fjellvag1986-of}, it is unclear whether the Te-Au bond remains a perturbation of the freestanding monolayer or competes with the intralayer Zr-Te network.
We address this question here, combining scanning tunneling microscopy (STM) on gold-exfoliated ZrTe$_5$ with \emph{ab initio} calculations of the ZrTe$_5$/Au(111) interface.
We show that the Au-Te interaction ruptures Te-Te bonds linking the ZrTe$_3$ chains, strongly disordering the ZrTe$_5$ monolayer on gold.
The Zr-Te bonds themselves survive: the mean bond length is unchanged, $2.975 \pm 0.024$~\AA{} freestanding against $2.982 \pm 0.048$~\AA{} on gold, while the spread doubles.

\section{Results and Discussion}

\begin{figure}[ht!]
\begin{center}
	\includegraphics[width = 1 \textwidth]{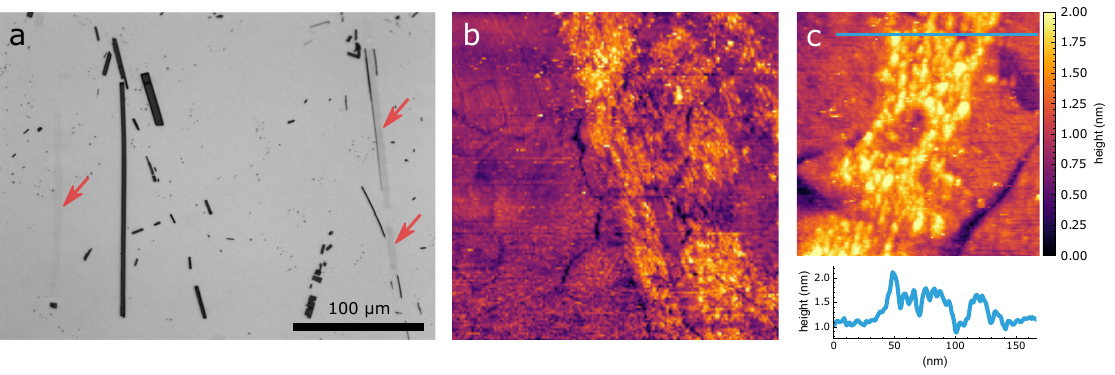}
	\caption{
		\textbf{Exfoliation of ZrTe$_5$ onto a gold substrate.}
		\textbf{(a)} Optical microscopy image of exfoliated ZrTe$_5$ crystals on a gold surface.
		The thinnest crystals on the surface are marked by red arrows.
		\textbf{(b)} STM topographic image in a representative thin area, as marked by red arrows in a).
		The image shows a disordered structure on the substrate.
		STM image measured at 300~K, setpoint: 50~pA, sample bias: 0.5~V.
		\textbf{(c)} Zoomed-in STM image of the disordered structure.
		Line section in the inset measured along the blue line.
		STM image measured at 9~K, setpoint: 500~pA, sample bias: 0.5~V.
		}
\label{fig:intro}
\end{center}
\end{figure}

We prepare the gold substrates by evaporation onto mica or Si wafers (see Methods); evaporation onto mica yields predominantly (111) facets~\cite{Dishner1998-au}.
The gold film is then peeled off the mica~\cite{Magda2015-nn}, exposing the pristine surface that had faced the mica; ZrTe$_5$ is exfoliated onto this freshly exposed surface immediately after stripping (see Methods).
Tellurides~\cite{Mirabelli2016-te} and 2D monolayers~\cite{Femi-Oyetoro2021-vs} are generally prone to ambient degradation, so we conduct exfoliation and gold stripping in a nitrogen glove box and transfer samples into the STM load lock through a vacuum suitcase.
From the load lock the samples are moved into the UHV chamber of the STM.
An optical microscopy image of a representative exfoliation is shown in Fig.~\ref{fig:intro}a, where the faintest optical contrast corresponds to monolayer flakes, marked by red arrows.
Representative STM measurements of these faint regions at 300~K and 9~K are shown in Fig.~\ref{fig:intro}b and c.
Low-magnification STM images reveal a disordered structure, with the ZrTe$_5$ layer discontinuous on the surface.
Multilayer flakes on the same samples show a crystalline surface, consistent with our earlier work on glove-box-prepared few-layer ZrTe$_5$~\cite{Tajkov2022-ur}, and indicate that the monolayer disorder is not caused by oxidation during exfoliation or transfer.
Control exfoliations of MoS$_2$, MoSe$_2$, WSe$_2$, and WTe$_2$ with the identical protocol yield crystalline monolayers with atomically resolved STM images (Supplementary Section~S4), ruling out the exfoliation procedure itself as the source of the disorder.

\begin{figure}[ht!]
\begin{center}
	\includegraphics[width = 0.7 \textwidth]{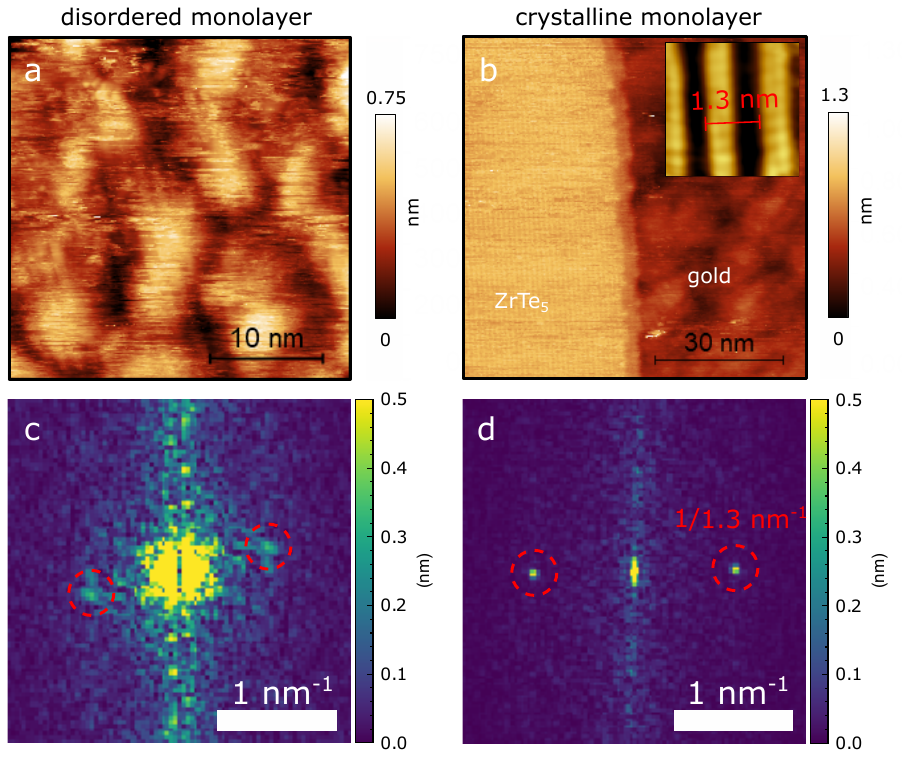}
	\caption{
		\textbf{STM measurements on gold-exfoliated ZrTe$_5$ monolayers.}
		\textbf{(a)} STM topographic image in the thinnest area, as marked by red arrows in Fig.~\ref{fig:intro}a.
		The image shows a disordered structure.
		\textbf{(b)} STM topographic image of a rare crystalline monolayer ZrTe$_5$ flake; its 0.6~nm apparent height above the gold identifies it as a single layer (Supplementary Section~S3).
		The high crystallinity is shown by the Fourier transform in d.
		Inset: atomic resolution STM image of the surface of the same flake; the red bar marks the unit cell period in the \emph{c} direction.
		\textbf{(c)} Fourier transform image of the data in panel a.
		\textbf{(d)} Fourier transform of the ZrTe$_5$ area of b.
		Red dashed circles mark the Fourier component of the periodicity in the \emph{c} direction.
		}
\label{fig:stm}
\end{center}
\end{figure}

We then navigate the STM tip optically to the monolayer candidate regions and attempt atomic resolution imaging at 9~K (Fig.~\ref{fig:stm}).
On most of the faintest flakes the topography remains highly disordered and the characteristic ZrTe chains are not resolved (Fig.~\ref{fig:stm}a).
In some monolayer regions the Fourier transform (Fig.~\ref{fig:stm}c) still shows a faint trace of the 1.3~nm Zr-chain periodicity, indicating that the disordered topography retains some residual structure of the ZrTe$_5$ lattice.
This quasi-ordered signature appears in roughly 10\% of the monolayer regions measured.
A possible origin is a locally weaker Te-Au coupling: the stripped gold exposes predominantly, but not exclusively, (111) facets~\cite{Dishner1998-au}, with STM showing areas of significant step formation, while our calculations address the flat (111) surface only, leaving the coupling strength on other facets an open question.
Across tens of samples the overwhelming majority of monolayer candidates show this disordered character.
In rare cases, however, a crystalline monolayer survives: the flake in Fig.~\ref{fig:stm}b resolves the quasi-one-dimensional ZrTe chains clearly, with the chain periodicity appearing as sharp peaks in the Fourier transform of the atomic-resolution image (Fig.~\ref{fig:stm}d).
Its apparent height of 0.6~nm above the gold identifies it as a single layer.
Apparent heights on a metal substrate fall below the crystallographic thickness, since the tunneling current mixes in the local density of states, orders of magnitude lower in ZrTe$_5$ than in gold, and since the chemisorbed contact layer is pulled in toward the surface.
Accounting for both effects, the disordered monolayer appears at 0.47~nm, the crystalline flake at 0.6~nm, and a bilayer would appear at roughly 1.2~nm, twice the measured value (Supplementary Section~S3).
The van der Waals-like spacing of this surviving monolayer points to a locally weaker Te-Au coupling, likely due to a densely stepped gold surface in this region, making Te-Au interaction weaker.
Thicker flakes on the same substrates invariably permit atomic resolution, consistent with our earlier work on few-layer ZrTe$_5$~\cite{Tajkov2022-ur}.

\begin{figure}[ht!]
\begin{center}
	\includegraphics[width = 1 \textwidth]{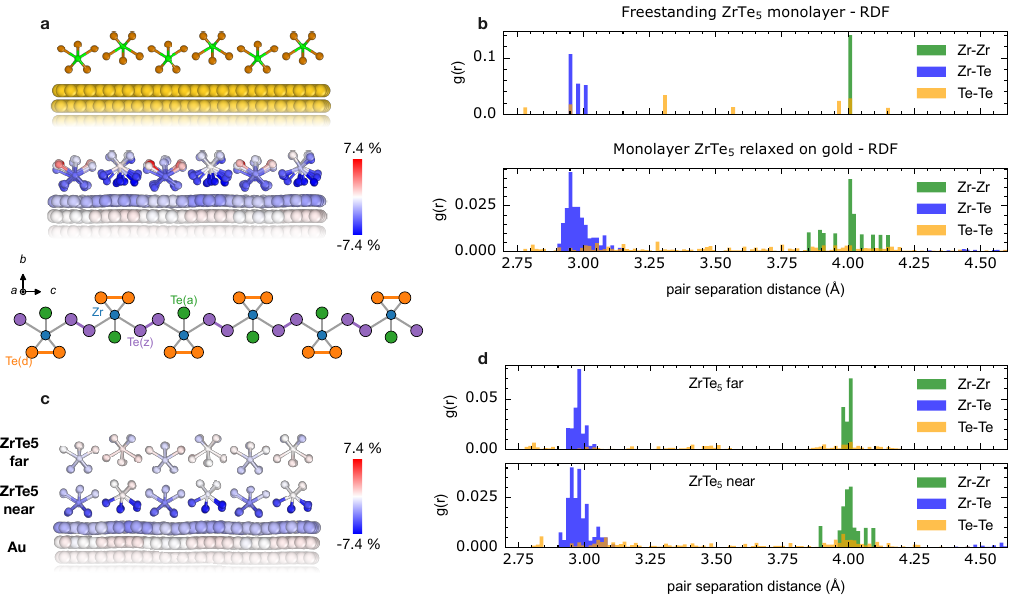}
	\caption{
		\textbf{\emph{ab initio} calculations of ZrTe$_5$ on gold.}
		Distribution of atomic distances in the few-layer ZrTe$_5$ crystals on the gold (111) surface in different configurations as calculated by VASP.
		\textbf{(a)} ZrTe$_5$ monolayer on top of gold (111), before \emph{ab initio} structural relaxation (top).
		Same monolayer after structural relaxation on top of gold (bottom).
		Inset at the bottom: schematic of the ZrTe$_5$ structure along the Zr chains, showing the Te sites: dimer (d), zigzag (z), apical (a), following the nomenclature of ref.~\cite{weng2014transition}.
		The site colors match those of the Bader analysis in Supplementary Section~S2.
		\textbf{(b)} Radial distribution function (RDF) for atom pairs in the freestanding and the relaxed monolayer on gold.
		\textbf{(c)} Bilayer ZrTe$_5$ on a gold surface after structural relaxation.
		The color scale in panels (a) and (c) shows the atomic strain calculated as the average of the diagonal strain tensor elements for each atom.
		The color scale indicates that the gold surface is itself perturbed by the presence of the ZrTe$_5$.
		\textbf{(d)} RDF of the two layers of the ZrTe$_5$ bilayer, closest to and farthest from the Au surface.
		The layer near the gold is much more disordered than the layer farthest from the support.
		}
\label{fig:atomic_pos}
\end{center}
\end{figure}

To understand the origin of the disorder, we turn to \emph{ab initio} calculations of the interface between the gold (111) surface and ZrTe$_5$ using the VASP code (see Methods).
Relaxed geometries of a monolayer and a bilayer on the gold slab are shown in Fig.~\ref{fig:atomic_pos}a and c; the monolayer becomes highly distorted on gold, while the top layer of the bilayer remains close to the freestanding geometry.

We quantify the distortion using the radial distribution functions (RDF) of the atomic distances.
In Fig.~\ref{fig:atomic_pos}b, top panel, we show the RDF of the relaxed freestanding monolayer as a reference.
ZrTe$_5$ has three inequivalent Te sites~\cite{osti_1277414,weng2014transition} (inset of Fig.~\ref{fig:atomic_pos}a): the zigzag Te(z), bonded to a single Zr atom in a distorted single-bond geometry; the dimer Te(d), bonded to two equivalent Zr atoms in a distorted L-shape; and the apical Te(a), bonded to two equivalent Zr atoms in a 2-coordinate geometry, with Zr-Te bond lengths close to 3~\AA\ that produce the three Zr-Te peaks in the freestanding RDF (Fig.~\ref{fig:atomic_pos}b).
After relaxation on the Au(111) surface, the Zr-Te distances broaden substantially: the mean bond length is nearly unchanged, $2.975 \pm 0.024$~\AA{} freestanding against $2.982 \pm 0.048$~\AA{} on gold, but the spread doubles and the range widens from 0.102 to 0.229~\AA{}.
The atomic arrangement of the ZrTe$_5$ layer thus becomes strongly disordered within the calculation cell (Supplementary Section~S1).
The broadening is most pronounced for the Zr-Te peaks, and also visible in the Zr-Zr and Te-Te distances.
In Fig.~\ref{fig:atomic_pos}a we visualize the distortion via the per-atom volumetric strain, calculated by the Ovito software~\cite{ovito} as the average of the diagonal strain tensor elements.
The gold surface atoms are distorted as well, as the ZrTe$_5$ layer pulls in toward the surface during relaxation, consistent with the strong Te-Au bonding quantified below.

The quantitative structural analysis of Supplementary Section~S1 resolves which bonds are more perturbed.
Every one of the 14 Zr atoms keeps exactly 8 Te neighbours upon adsorption, so the bicapped trigonal prismatic units of the ZrTe$_3$ chains are distorted but not broken up.
The Te-Te network is what yields: 14 of the 32 short Te-Te contacts of the freestanding layer stretch beyond 3.0~\AA{} on gold, and a quarter of all Te-Te pairs move into distance windows that are empty in the crystalline layer.
The seven discrete Te-Zr-Te bond-angle groups broaden from an average width of 1.9$^\circ$ to 5.4$^\circ$ and merge.
Thus, the coordination polyhedra are scrambled internally rather than sheared in one direction.
The cause of the disorder is the interface with the gold: 23 of the 24 Te atoms facing the gold form Te-Au bonds with a mean length of 2.89~\AA{}, far below the van der Waals contact distance of 3.72~\AA{}, so the exposed Te plane is chemisorbed on the gold.
The Au-Te chemisorption thus ruptures the weak Te-Te bonding while the ZrTe$_3$ chains persist as distorted structural units.
Bulk thermochemistry supports this: the standard formation enthalpy of ZrTe$_2$, $-294.1 \pm 6.7$~kJ/mol per formula unit~\cite{Johnson1985-zr}, exceeds that of AuTe$_2$, $-18.6 \pm 3.0$~kJ/mol~\cite{Andon1971-te}, by an order of magnitude, so the interface bond can out-compete the weak Te-Te links but not the Zr-Te bonds of the chains.

The charge redistribution at the interface confirms the covalent character of the Te-Au bond (Supplementary Section~S2).
The charge-density-difference maps show electron depletion around the interface Te and Au atoms, with accumulation in the bonding region between them.
Bader analysis quantifies the transfer: the Te atoms in contact with the gold lose up to 0.35 electrons each, while the net transfer to the gold is modest, 0.37 electrons per unit cell, a hole doping of $6.6\times10^{13}$~cm$^{-2}$.
The Bader charge of the Zr atoms changes by less than 0.05~e.
The perturbation decays within roughly 2~\AA{} of the contact, and the dimer Te of the top surface recover their freestanding charges.

In the bilayer, the layer in direct contact with the gold (labelled ``near'' in Fig.~\ref{fig:atomic_pos}d) shows an RDF distorted similarly to the monolayer case, while the layer farthest from the gold (``far'') has a less disordered RDF (Fig.~\ref{fig:atomic_pos}d, top panel).
Across the three cases, the distortion decreases monotonically with distance from the gold: the monolayer is the most distorted, the bilayer surface closest to the metal is equally perturbed, and the outer bilayer surface starts approaching the freestanding lattice.
The hole doping follows this trend: less than 5\% of the doping of the layer in contact with the gold reaches the second layer (Supplementary Section~S2).
This supports the STM observation that the top surface of few-layer flakes is crystalline.
Whether the electronic structure of the top layer in the distorted bilayer approximates that of a freestanding monolayer remains an open question for direct spectroscopic measurements.

To quantify the interaction strength, we compute the bonding energies for different stackings (see Methods).
The bonding energy between the gold slab and a monolayer of ZrTe$_5$ is 2.43~eV per unit cell, more than double the 1.10~eV per unit cell computed for the interlayer bond between a five-layer ZrTe$_5$ stack and an adjacent monolayer.
Expressed per unit area (55.1~\AA$^2$ per unit cell), the adhesion of 44~meV/\AA$^2$ is 2.2 times the interlayer cohesion of 20~meV/\AA$^2$; for MoS$_2$ on Au(111) the calculated ratio of adhesion to interlayer cohesion is only 1.2--1.5~\cite{Velicky2018-wj,Huang2020-tw}.
The energy required to separate the top layer from a gold-supported bilayer is 1.16~eV per unit cell.
These energies show that gold contact preferentially peels a single ZrTe$_5$ layer from the bulk; the price paid by the extracted monolayer is the strong local distortion documented above.
These numbers quantify the local distortion within the periodic 14-formula-unit supercell; the loss of long-range order itself is established by the STM measurements of Fig.~\ref{fig:stm}.
Since the relaxed geometry is a zero-temperature local minimum reached from a crystalline starting configuration, it sets a lower bound on the disorder the Au-Te interaction can produce.

Finally, we return to the possibility of the quantum spin Hall state~\cite{weng2014transition} surviving in the disordered ZrTe$_5$.
Our results show that the gold-supported monolayer is not a likely platform for this phase.
The chemisorption ties the layer to a metal that hybridizes with the interface Te and hole-dopes the layer (Supplementary Section~S2).
The induced disorder likely scrambles a band inversion that strains below 1\% strongly influence~\cite{Tajkov2022-ur,mutch2019evidence}.
Either effect alone precludes the quantum spin Hall state.
Accessing the topological phase of monolayer ZrTe$_5$ therefore requires weakly coupled supports, as demonstrated by Al$_2$O$_3$-assisted exfoliation with hBN encapsulation~\cite{Zhuo2022-ae} and by molecular beam epitaxy on SiC/graphene~\cite{Xu2024-hv}.

\section{Conclusions}

The disorder of the ZrTe$_5$ monolayers arises from the combination of two ingredients: the quasi-covalent Te-Au bond~\cite{Zhang2023-mf,Asikainen2025-xf,Bouaziz2020-nx}, which is generic to Te-terminated crystals on gold, and the quasi-one-dimensional bonding network of ZrTe$_5$, whose weak inter-chain Te-Te links the interface bond out-competes.
The Te-Au bond alone does not destroy monolayer order: MoTe$_2$, WTe$_2$, PtTe$_2$, and PdTe$_2$ exfoliate into crystalline monolayers on gold~\cite{Velicky2018-wj,Huang2020-tw}.
Milder local distortions of the contact layer may nevertheless be present in these systems, and crystals with weakly cross-linked, anisotropic layers, such as the sibling pentatelluride HfTe$_5$ or the chain compound ZrTe$_3$, are the natural candidates for ZrTe$_5$-like disorder.
Conversely, the disordered ZrTe$_5$ monolayer may itself be of interest in the context of atomically thin amorphous and disordered materials; the synthesis of monolayer amorphous carbon~\cite{Toh2020-ib} and of nanodroplet-grown amorphous metal chalcogenides~\cite{Shi2025-gy} shows that disorder at the 2D limit can serve as a useful degree of freedom for tuning conductivity, work function, and catalytic activity in ultrathin films.
The rare crystalline monolayers that survive exfoliation, together with the possible facet dependence of the quasi-ordered areas, suggest that engineering the gold surface toward weaker Te-Au coupling may yet yield intact monolayer ZrTe$_5$ for probing the predicted 2D topological insulator phase.

\section{Methods}
\label{methods}
The optimized geometry and electronic properties of the crystal were obtained using density functional theory (DFT) as implemented in VASP~\cite{kresse_1993,kresse_1996}.

\textbf{ZrTe$_5$ exfoliation}

Samples were prepared from ZrTe$_5$ crystals purchased from HQ Graphene, inside a nitrogen-filled glove box with a water and oxygen content of 0.1~ppm or better.
The substrates were template-stripped (``stripped gold'') surfaces~\cite{Magda2015-nn}, prepared from gold films evaporated onto mica or Si wafers.
Si wafer supports were glued onto the evaporated gold films with epoxy and, after curing, the substrates were loaded into the glove box.
Inside the glove box, the growth substrate was stripped off and ZrTe$_5$ flakes were exfoliated immediately onto the freshly exposed gold surface.
No ultrasonication was used at any step of the procedure.
Both types of gold film produce the same outcome, monolayer ZrTe$_5$ becomes disordered, and the result is not affected by flame annealing the gold before exfoliation.
The samples were inspected by optical microscopy to identify monolayer candidates, mounted onto an STM sample holder, and transferred to the load lock of the STM chamber in a vacuum suitcase.
The same procedure was followed for all TMDC flakes shown in the Supplementary Information.

\textbf{STM measurements}

STM measurements were performed on an RHK PanScan Freedom microscope operating at 300~K and 9~K in ultra-high vacuum with a base pressure of $5\times10^{-11}$~Torr.
STM tips were mechanically cut from Pt/Ir (90/10) wire.
The tip is positioned above a selected flake in two steps.
Before loading, the sample is mapped with a laboratory optical microscope (Fig.~\ref{fig:intro}a), and the position of each monolayer candidate is recorded relative to the thick, high-contrast flakes and the sample edges, which serve as landmarks.
The PanScan Freedom has an open body with direct line of sight to the tip-sample junction, which we view through a viewport with a long-working-distance zoom microscope and a camera.
The thick flakes and the grain structure of the gold film are visible in this camera image, while the monolayer regions are not.
The gold film is mirror-like, so the tip and its reflection are both visible; this shows the tip-sample separation during the coarse approach and places the tip apex laterally to within roughly 5~$\upmu$m.
Using the thick flakes as landmarks, the sample is moved under the tip with the coarse piezo motors until the apex sits over the target region, and the tip is approached.
Since this lateral uncertainty is comparable to the width of a monolayer ribbon, the final step relies on the STM itself: we record micrometer-scale topographic images and step the sample with the coarse motors in sub-micrometer increments until the edge of the flake is found, identified by its step height above the atomically flat gold terraces.
The elongated shape of the ZrTe$_5$ ribbons, tens of micrometers long along the chain direction, makes this search efficient.

\textbf{VASP}

Calculations used projector-augmented wave pseudopotentials and the optB86b-vdW functional~\cite{klimes_2011}, which includes a non-local correlation correction that approximately accounts for dispersion interactions.
We have found that this functional produces the most accurate geometries in comparison with experiment.
The plane-wave cutoff energy was set to $500\,\rm eV$ in all calculations.
Structural optimizations used a $20\times 6\times 1$ Monkhorst-Pack $k$-point set and were run until all atomic forces fell below $3\,\rm meV/$\AA.
All structural relaxations and binding-energy calculations were carried out without spin--orbit coupling (SOC), following a widely used approach in density-functional studies~\cite{martin2020,weng2014transition}.

\textbf{VASP calculations of ZrTe$_5$ on gold}

We modelled the gold (111) surface as a 6-layer slab with the bottom 2 layers fixed during optimization.
To obtain a supercell commensurate with ZrTe$_5$, we used 7 unit cells of the crystal.
The supercell unit vectors are $\mathbf{A}_1 = -3\mathbf{a}+\mathbf{c}$ and $\mathbf{A}_2 = 4\mathbf{a}+\mathbf{c}$, where $\mathbf{a}=\left(a,0,0 \right)$ and $\mathbf{c}=\left(0,0,c \right)$.
This gives a $27.192^\circ$ rotation between the gold and the ZrTe$_5$ unit cell vectors and a small ($<2\%$) lattice mismatch between the two slabs.
The combined gold slab and monolayer ZrTe$_5$ supercell contains 312 gold atoms, 14 Zr atoms, and 70 Te atoms.
These calculations were performed using a $2\times 2\times 1$ Monkhorst-Pack $k$-point set, which is sufficient for a supercell of this size.

\section*{Competing interests}
The authors declare no competing interests.

\section*{Data Availability}
The datasets generated during and/or analysed during the current study are available in the Zenodo repository, \href{https://doi.org/10.5281/zenodo.19680005}{doi.org/10.5281/zenodo.19680005}.

\section*{Funding}
The work was conducted within the framework of the MTA - HUN-REN EK Lendület ``Momentum'' Topology in Nanomaterials Research Group through project LP2024-17.
Financial support from NKFIH through grants \'{E}lvonal KKP 138144, Excellence 151372, K146156 and TKP2021-NKTA-05 is also acknowledged.
ZT acknowledges support from the J{\'a}nos Bolyai Research Scholarship of the Hungarian Academy of Sciences. This project is supported by the TRILMAX Horizon Europe consortium (Grant No. 101159646).
We acknowledge the Digital Government Development and Project Management Ltd. for awarding us access to the Komondor HPC facility based in Hungary.

\bibliographystyle{unsrtnat}
\bibliography{zrte5-mono-gold}

\section*{Author contributions}
KK prepared the samples and performed the STM measurements under the supervision of PNI.
ZT and JK performed the calculations.
LT contributed to data analysis and interpretation.
PNI conceived and coordinated the project, with assistance from JK.
PNI and ZT wrote the manuscript with input from all authors.

\end{document}